\RequirePackage{lineno}

\documentclass[aps, preprint, nofootinbib, superscriptaddress, showpacs]{revtex4-1}
\usepackage{subfigure}
\usepackage{graphicx}
\usepackage{amsmath}
\usepackage{cleveref}
\usepackage{color}
\usepackage[todos]{CMImacros}

\DeclareMathOperator{\somb}{somb}

\begin{document}

\title{Superresolving second-order correlation imaging using synthesized colored noise speckles}
\author{Zheng Li$^\ddagger$}
\email{Email: zheng.li@pku.edu.cn}
\affiliation{State key Laboratory for Mesoscopic Physics, School of Physics, Peking University, Beijing 100871, China}
\author{Xiaoyu Nie$^\ddagger$}
\affiliation{Department of Physics and Astronomy, Texas A\&M University, College Station, Texas, 77843, USA}
\affiliation{Physics Department, Xi'an Jiaotong University, Xi'an, Shaanxi 710049, China}
\author{Fan Yang$^\ddagger$}
\affiliation{Department of Physics and Astronomy, Texas A\&M University, College Station, Texas, 77843, USA}
\affiliation{Physics Department, Xi'an Jiaotong University, Xi'an, Shaanxi 710049, China}
\author{Xiangpei Liu}
\affiliation{Department of Physics and Astronomy, Texas A\&M University, College Station, Texas, 77843, USA}
\affiliation{Hefei National Laboratory for Physics Science at the the Microscale and Department of Modern Physics, University of Science and Technology of China, Hefei, Anhui 230026, China}
\author{Dongyu Liu}
\affiliation{State key Laboratory for Mesoscopic Physics, School of Physics, Peking University, Beijing 100871, China}
\author{Xiaolong Dong}
\affiliation{State key Laboratory for Mesoscopic Physics, School of Physics, Peking University, Beijing 100871, China}
\author{Xingchen Zhao}
\affiliation{Department of Physics and Astronomy, Texas A\&M University, College Station, Texas, 77843, USA}
\author{Tao Peng}
\email{Correspondence to: Tao Peng, Email: taopeng@tamu.edu}
\affiliation{Department of Physics and Astronomy, Texas A\&M University, College Station, Texas, 77843, USA}
\author{M. Suhail Zubairy}
\affiliation{Department of Physics and Astronomy, Texas A\&M University, College Station, Texas, 77843, USA}
\author{Marlan O. Scully}
\affiliation{Department of Physics and Astronomy, Texas A\&M University, College Station, Texas, 77843, USA}
\affiliation{Baylor Research and Innovation Collaborative, Baylor University, Waco, 76706, USA}
\affiliation{Princeton University, Princeton, New Jersey 08544, USA}

\footnote{$^\ddagger$ These authors contributed to this work equally.}


\setpagewiselinenumbers
\modulolinenumbers[1]

\definecolor{oldtxtcolor}{rgb}{0.00, 0.0, 0.5}
\definecolor{newtxtcolor}{rgb}{0.00, 0.3867, 0.00}
\definecolor{newtxtcolor}{rgb}{0.00, 0.0, 1}
\definecolor{oldtxtcolor}{rgb}{1.00, 0.0, 0.00}

\def\verX{12}
\def\verO{1}
\def\verN{2}
\def\verON{12}

\ifx\verX\verO
 \newcommand { \oldtxt }[1] {{\color{oldtxtcolor}{#1}}}
 \newcommand { \newtxt }[1] {}
\fi
\ifx\verX\verN
 \newcommand { \oldtxt }[1] {}
 \newcommand { \newtxt }[1] {{\color{newtxtcolor}{#1}}}
\fi
\ifx\verX\verON
 \newcommand { \oldtxt }[1] {{\color{oldtxtcolor}{#1}}}
 \newcommand { \newtxt }[1] {{\color{newtxtcolor}{#1}}}
\fi

\maketitle

\textbf{
We present a novel method to synthesize non-trivial speckles that can enable superresolving second-order correlation imaging. The speckles acquire a unique anti-correlation in the spatial intensity fluctuation by introducing the blue noise spectrum to the input light fields through amplitude modulation. Illuminating objects with the blue noise speckle patterns can lead to a sub-diffraction limit imaging system with a resolution more than three times higher than first-order imaging, which is comparable to the resolving power of ninth order correlation imaging with thermal light. Our method opens a new route towards non-trivial speckle generation by tailoring amplitudes of the input light fields and provides a versatile scheme for constructing superresolving imaging and microscopy systems without invoking complicated higher-order correlations.
}

\section{Introduction}
The Rayleigh diffraction limit specifies the minimum separation between two incoherent point sources that can be resolved into distinct objects~\cite{Abbe:1873aa,rayleigh1879xxxi}.
Over the decades, there has been vibrant research to develop superresolving imaging techniques that circumvent the Rayleigh limit by using quantum optical $N$-photon correlation~\cite{d2001two,li2008optical,boto2000quantum}, structured illumination~\cite{gustafsson2000surpassing,zeng2014nanometer, classen2017superresolution}, Rabi oscillation~\cite{kiffner2008resonant, liao2010quantum}, fluorescence saturation~\cite{hell1994breaking}, photoswitching~\cite{hess2006ultra,rust2006sub,betzig2006imaging}, and higher-order detection of classical light~\cite{hemmer2006quantum,guerrieri2010sub}.
For the case of classical light illumination, the resolution can be improved by a factor of $\sqrt{2}$ with second-order speckle correlation of thermal light~\cite{oh2013sub}, and can be further enhanced by spatial filtering~\cite{sprigg2016super,chen2017sub}.
Higher $N$th-order correlation provides further enhancement of resolution by $\sqrt{N}$ times~\cite{zhou2012resolution,oppel2014directional,zhang2016high}.

As a fundamental component of the modern optics toolbox, laser speckles appear when coherent light impinges upon a scattering sample and are typically generated by modulating a laser beam with rotating ground glass~\cite{martienssen1964coherence}, or spatial light modulator (SLM)~\cite{shapiro2008computational}.
The amplitudes of the speckles are distributed following the Rayleigh statistics, resulting in a negative exponential intensity probability density function (PDF)~\cite{goodman2005introduction}, known as pseudo-thermal light~\cite{chen2013100,li2020photon}.
Recently, interest has been aroused in tailoring and generating non-trivial speckle patterns, \textit{e.g.}, by phase modulation of the input light fields at the Fourier plane of SLM, speckles exhibiting non-Rayleigh statistics can be produced and used to optimize speckle illumination imaging~\cite{waller2012phase,takasaki2014phase,bromberg2014generating,bender2018customizing,kondakci2016sub,li2016generation,zhou2017superbunching,bender2018customizing}.

This work presents a novel method to synthesize speckles that possess non-trivial spatial correlation via amplitude modulation of the input light fields and achieve superresolving second-order correlation imaging with the obtained speckle illumination. 
The speckle patterns are generated by tailoring the light fields' amplitudes using a digital micromirror device (DMD). They have a blue noise power spectrum with spectral intensity increasing for higher spatial frequencies~\cite{ulichney1988dithering,ulichney1988dithering}. In addition, the PDF of the blue noise speckles is made to obey the Rayleigh statistics, which is the same as the white noise speckles.
The unique property of our blue noise speckles is the negative correlation in intensity fluctuation between neighboring spatial pixels, which endows the second-order correlation imaging with superresolving power when the object is illuminated by such speckles.
We experimentally demonstrate that the resulting image shows $>3$ times higher resolution than the first-order imaging system, which essentially surpasses the $\sqrt{2}$ times enhancement by conventional second-order imaging using thermal light.

\section{Materials and Methods}
\subsection{Synthesized colored noise speckles and the spatial correlation}
\begin{figure*}
		\includegraphics[width=\linewidth]{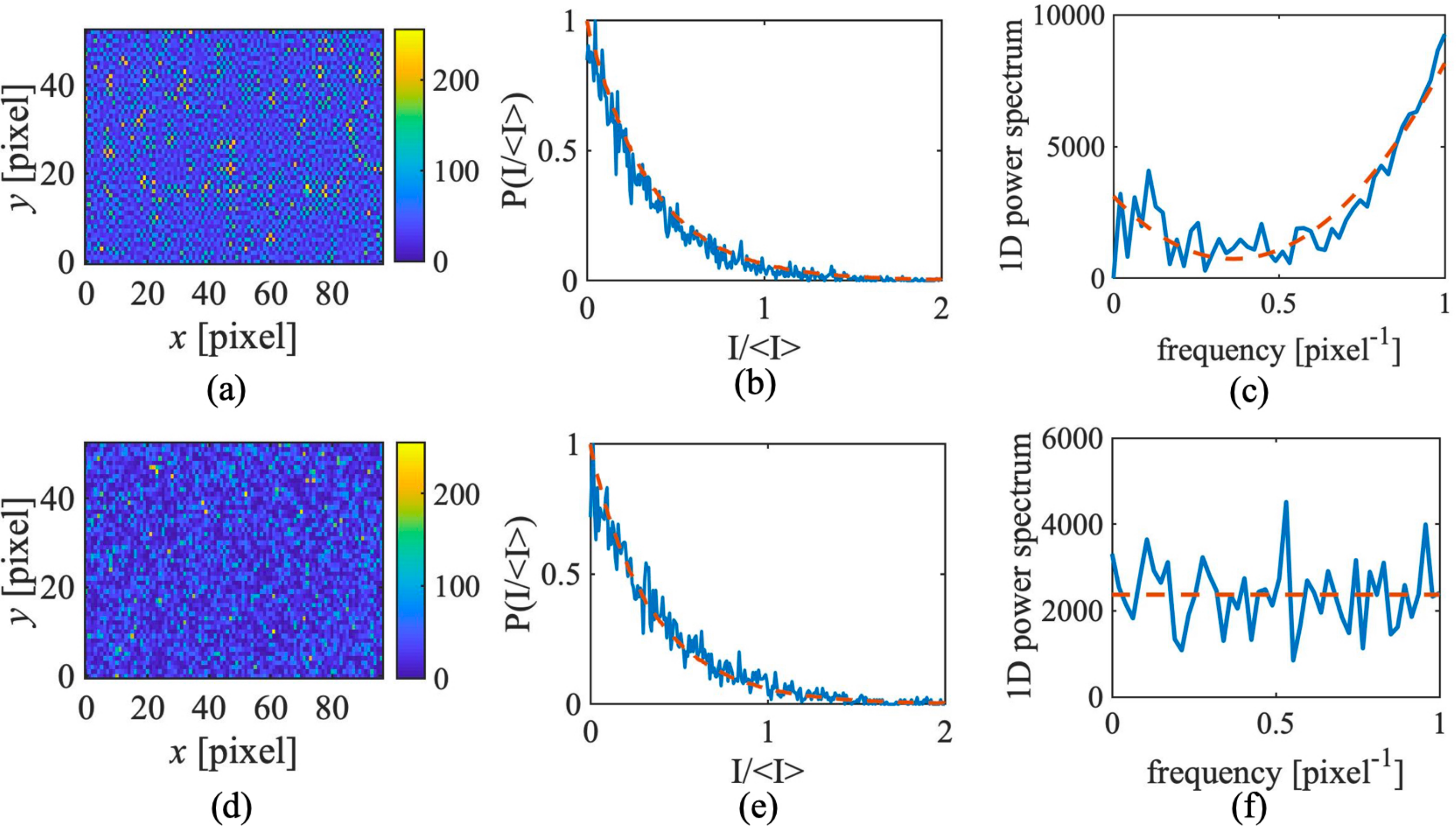}
\caption{\label{fig:pattern} %
(a) A typical 2D blue noise speckle pattern, (b) its probability distribution, and (c) its spectrum distribution.
(d) A typical 2D white noise speckle pattern, (e) its probability distribution, and (f) its spectrum distribution. Both the blue and white noise speckle patterns obey the same Rayleigh distribution. 
The red dashed lines in (b), (c), (e), and (f) are the corresponding plots based on averaging over all the 3000 blue and white noise patterns used in the experiment, respectively. The blue noise power spectrum scales asymptotically as $\omega^4$ for spatial frequency $\omega$.
}
\end{figure*}
The noise speckle patterns are generated and imprinted into the amplitude of the input light field by DMD.
We take the spectral power distribution of the noise speckles to be $I(\omega)\simeq C_0 \delta (\omega) + C \omega^n$ for spatial frequency $\omega$. White noise ($n=0$), and blue noise with $n=4$ are used in our experiment. After applying random conjugated symmetric phases to the power spectrum, an inverse Fourier transform is performed to obtain the noise speckle patterns in grayscale.
The generated patterns obey Gaussian statistics with low spatial correlation, which makes the imaging system suffer from measurement noise.
Thus we redistribute the speckle PDF to satisfy Rayleigh statistics with the local intensity transformation~\cite{Bender:19}
\begin{align}\label{PDF}
\int^{I_0}_0 P(I) d I =\int^{I'_0}_0 P(I') d I',
\end{align}
where $P(I')$ is the PDF of the original created Gaussian patterns. The patterns are then one-to-one mapped to the negative Rayleigh distribution
\begin{align}\label{PDF_Rayleigh}
P(I)=(1/\bar{I})\exp{(-I/\bar{I})}
\end{align}
The resulting Rayleigh blue noise speckle pattern, its PDF, and its power spectrum are shown in Figs.~\ref{fig:pattern}(a), (b), and (c), respectively.
The white noise Rayleigh speckle pattern, its PDF, and power spectrum are also shown in Figs.~\ref{fig:pattern}(d), (e), and (f), respectively. Unlike previous works that obtain non-trivial second-order correlations via tailoring the speckles' statistics, we show that the blue noise speckle has Rayleigh statistics resulting in the same autocorrelation for each pixel, same as the white noise speckle. However, a non-trivial correlation between one pixel and its neighboring pixel appears due to the modified spectrum distribution. 

\begin{figure*}[!hbt]
\includegraphics[width=\linewidth]{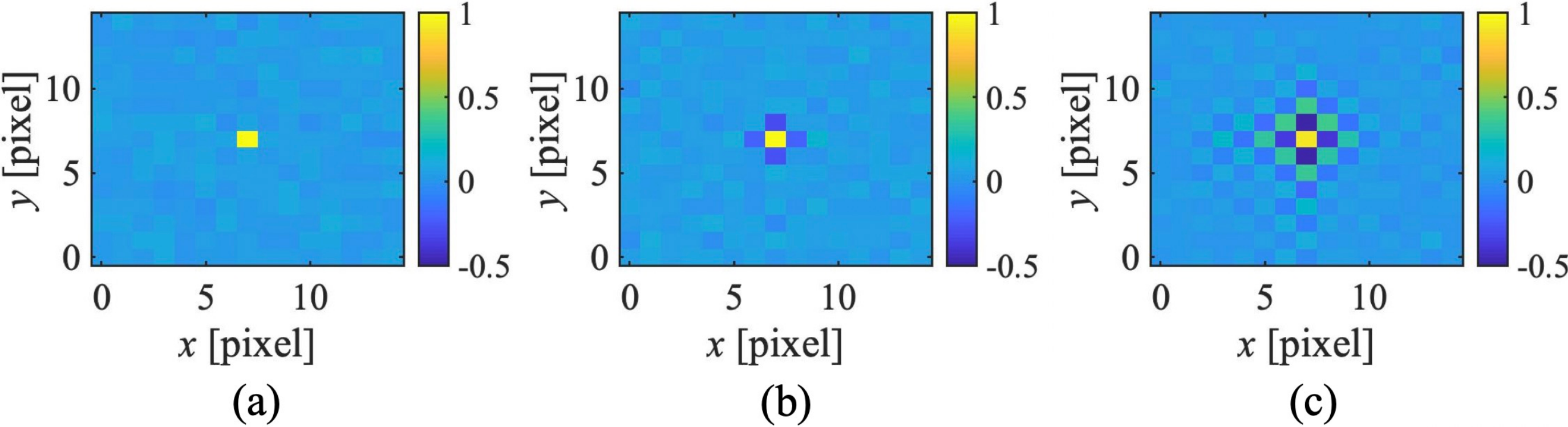}
\caption{2D spatial intensity fluctuation correlation of speckle patterns for an arbitrary pixel and its neighboring pixels.   (a) white noise, (b) blue noise ($n=1$), and (c) blue noise ($n=4$). It is shown that the white noise has a zero correlation between different pixels, as expected. For blue noise, there exists a significantly negative spatial correlation between neighboring pixels; the fluctuation correlation between adjacent pixels is $-0.21$ for $n=1$ blue noise speckles and reaches $-0.44$ for $n=4$ blue noise speckles.}
   \label{fig:experiment_theory}
\end{figure*}

We then examine the spatial correlation of the white noise and blue noise Rayleigh speckles.
To simplify the calculation without loss of generality, we start with one dimension. 
The spatial intensity fluctuation correlation, which is central to the image formation, is defined as  
\begin{align}\label{fluctuation_correlation_defination}
\gamma^{(2)} (\tilde{x}) =\langle \Delta I(x_1) \Delta I(x_2)\rangle = \frac{\langle I(x_1)I(x_2)\rangle}{\langle I(x_1) \rangle\langle I(x_2)\rangle}-1,
\end{align}
where $\tilde{x}=x_2 - x_1$.
Given the power spectrum density of the patterns, the intensity fluctuation correlation is 
\begin{align}\label{intensity_flutuation_correlation}
\gamma^{(2)} (\tilde{x})=\mathcal{F}^{-1}\{|C \omega^n|^2\}(\tilde{x}).
\end{align}

For white noise speckles, the intensity of each pixel fluctuates randomly and independently:
\begin{align}\label{white_fluctuation_correlation}
\gamma^{(2)}_{\mathrm{w}} (\tilde{x})=\mathcal{F}^{-1}\{|C_{\mathrm{w}}|^2\}(\tilde{x}) \sim \delta(\tilde{x}).
\end{align}
There is no correlation between adjacent pixels.

For blue noise speckles, we have the intensity fluctuation correlation as  
\begin{align}\label{blue_fluctuation_correlation}
\gamma^{(2)}_{\mathrm{b}} (\tilde{x})=\mathcal{F}^{-1}\{|C_{\mathrm{b}} \omega ^4|^2\}(\tilde{x}).
\end{align}
This results in a rather complicated form as compared with the white noise case. The Rayleigh blue noise speckles possess significantly negative spatial correlation of intensity fluctuation between neighboring pixels, with the slope and negativity of $\gamma^{(2)}_{\mathrm{b}}(\tilde{x})$ essentially enhanced for larger blue noise order $n$. The correlation oscillates as distance between two pixels increases, and reaches zero when they are well separated. We plot the 2D correlation functions of the intensity fluctuation of white noise speckles (Fig.~\ref{fig:experiment_theory} (a)), blue noise speckles for $n=1$ (Fig.~\ref{fig:experiment_theory}(b)), and $n=4$ (Fig.~\ref{fig:experiment_theory}(c)), using 3000 patterns for each noise type. The fluctuation correlation between adjacent pixels reaches $-0.44$ for $n=4$ blue noise speckles.
The striking anti-correlation of intensity fluctuation correlation readily provides the capability to resolve objects beyond the Rayleigh diffraction limit.

\subsection{Measurement method of second-order correlation imaging}
\begin{figure*}
\centering\includegraphics[width=0.55 \linewidth]{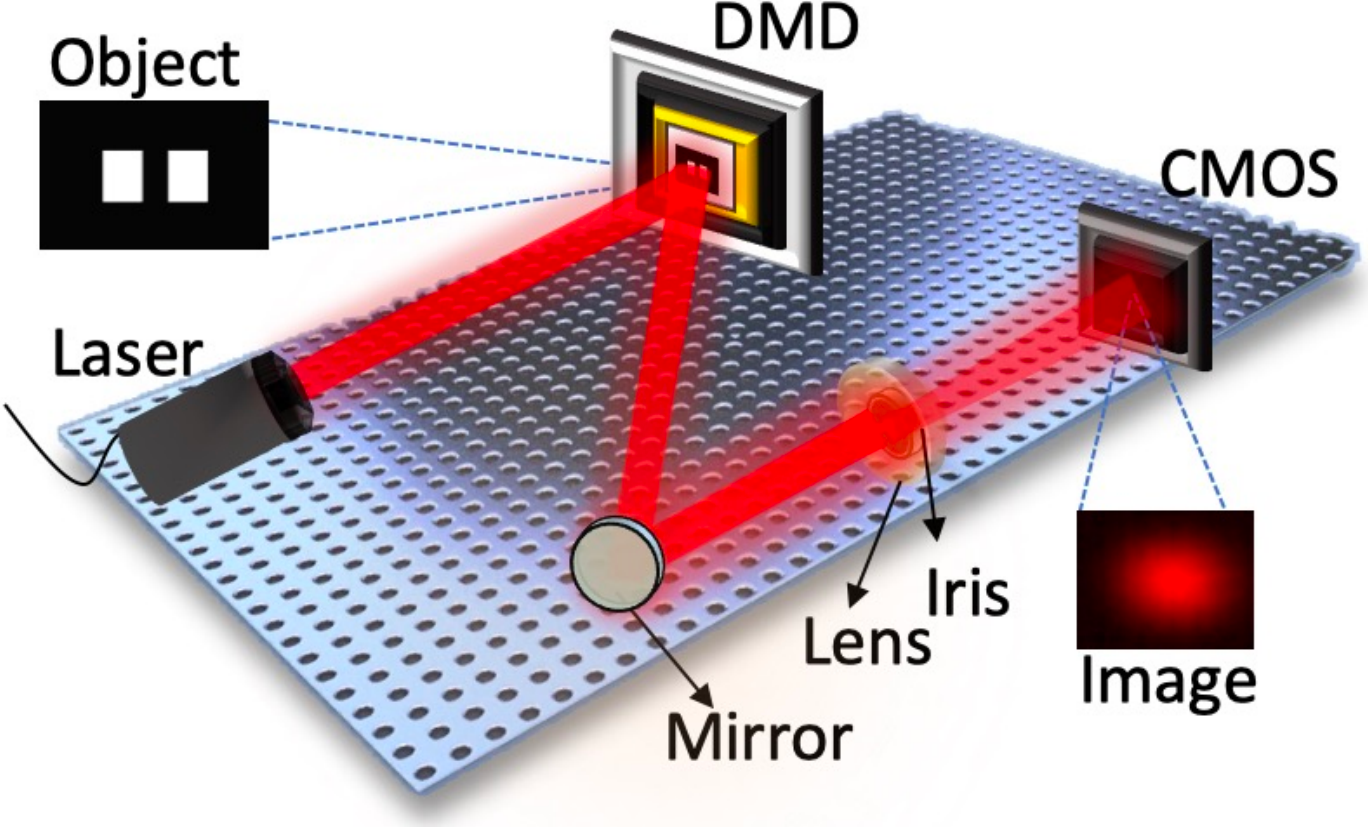}
\caption{\label{fig:setup} Schematic setup of the second-order correlation imaging system. A CW laser is reflected by a DMD. The synthesized speckle patterns and digital double-slit are loaded onto the DMD. The reflected light propagates to the imaging plane by an imaging lens. An 800 $\mu$m diameter iris pinhole is put right behind the lens to adjust the effective aperture of the lens. The CMOS sensor at the image plane records the intensity distribution $I(r)$ of the double-slit image.}
\end{figure*}

To demonstrate the corresponding resolving power of second-order correlation imaging with blue noise speckle illumination, we employ the experimental setup shown in Fig.~\ref{fig:setup}.
An input CW laser is reflected by the DMD, on which the speckle patterns of synthesized spatial spectrum and a digital object (double-slit pattern) are loaded.
The convoluted double-slit and speckle patterns are then imaged by the complementary metal oxide semiconductor (CMOS) sensor through an imaging lens.
The effective aperture of the imaging lens can be adjusted by a pinhole behind the lens.
The DMD contains pixelated mirrors with an area of 16 $\mu$m $\times$ 16 $\mu$m per pixel. The double-slit has a dimension of 4 pixels in height, 3 pixels in width for each slit, and 5 pixels of center-to-center separation. The noise pattern consists of 54 $\times$ 98 independent pixels (each pixel is 4 $\times$ 4 DMD pixels). The diameter of the pinhole is chosen to be 800 $\mu$m.

\section{Results and Discussion}
\subsection{Superresolution with blue noise speckle patterns}\label{sec3}
Due to existence of diffraction, the light reflected from each slit diffuses and can become overlapped with each other.
We measured the first-order image $I(r_{\mathrm{i}})$ and the second-order correlation image with white and blue noise ($n=4$) speckles illumination, which is constructed from the intensity fluctuation correlation of measured intensities on each pixel of the image plane.
The intensity distribution of the light field on the image plane can be expressed as
\begin{align}\label{first_order}
\langle I(r_{\mathrm{i}})\rangle \propto \left\langle\int\mathrm{d}r_{\mathrm{o}}  I(r_{\mathrm{o}})|T(r_{\mathrm{o}})|^2|h(r_{\mathrm{i}},r_{\mathrm{o}})|^2 \right\rangle,
\end{align}
where $I(r_{\mathrm{o}})$ is the intensity distribution of the speckle pattern and $r_{\mathrm{o}}$ is the coordinate on the object plane. $T(r_{\mathrm{o}})$ denotes the object aperture function, and \begin{equation}\label{PSF}
    h(r_{\mathrm{i}},r_{\mathrm{o}})\propto \somb \left(\frac{2\pi R |r_{\mathrm{o}} +r_{\mathrm{i}}/m|}{\lambda s_{\mathrm{o}}} \right) 
\end{equation}
is the point spread function (PSF)~\cite{shih2018introduction}. Here $R$ is the effective radius of the image lens,  $m=s_{\mathrm{i}}/s_{\mathrm{o}}$ is the magnification factor of the imaging system, $s_{\mathrm{o}}$ is the distance between object plane and lens, $s_{\mathrm{i}}$ is distance between lens and image plane, $\somb(r)=J_1(r)/r$, and $J_1(r)$ is the first-order Bessel function of the first kind.

Since the slit width is smaller than the noise speckle size, we can treat the two separated slits as two points in the object plane, at $r_1$ and $r_2$. The first-order image is then 
\begin{align}
\label{first_order_image}
\langle I(r_{\mathrm{i}})\rangle \propto h^2(r_{\mathrm{i}},r_1) +h^2(r_{\mathrm{i}},r_2)
\,.
\end{align}
The second-order image measured by illumination of white and blue noise speckle pattern is
\begin{align}
\label{second_order_image}
    \langle \Delta I_{\mathrm{w/b}}^2(r_{\mathrm{i}})\rangle 
    &\propto \left\langle\int \mathrm{d}r\mathrm{d}r_{\mathrm{o}}\,  \gamma^{(2)}_{\mathrm{w/b}}(r,r_{\mathrm{i}})|T(r_{\mathrm{o}})|^4 h^4(r,r_{\mathrm{o}}) \right\rangle
    \,.
\end{align}
In Eq.~(\ref{second_order_image}), the image formation depends strongly on the properties of spatial correlation of illumination light fields with different types of speckle patterns and cannot be expressed as the overlap of two PSFs from two separated points $r_1, r_2$. 
For the blue noise speckles, their significantly negative cross correlation of the pixelwise intensity fluctuation and the contribution of the autocorrelation from each pixel results in the superresolution of the formed image.
\begin{figure*}[hbt!]
\centering
\includegraphics[width=\linewidth]{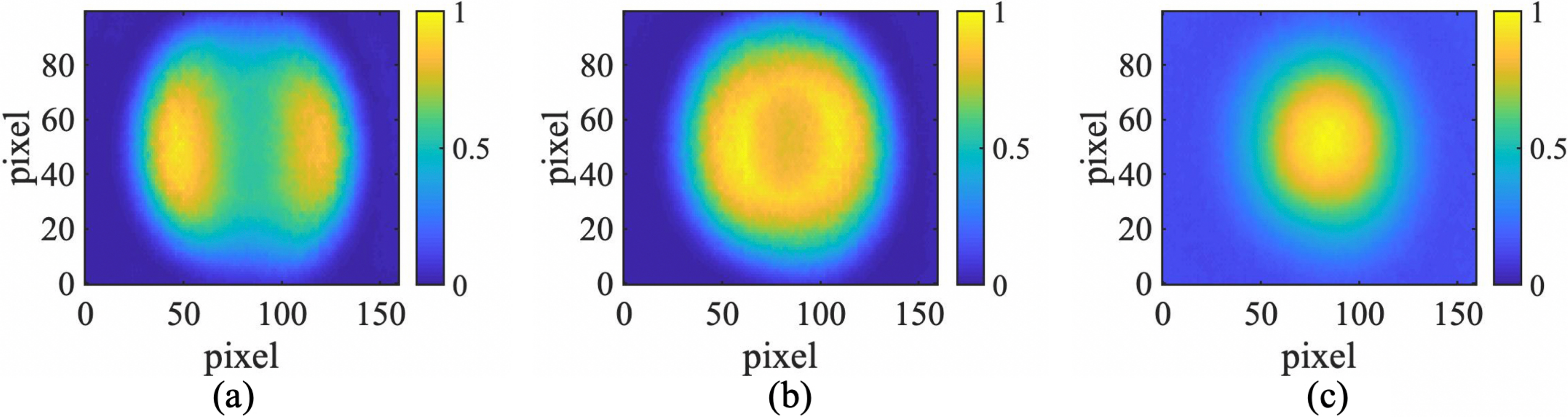}
	\caption{second-order image with Rayleigh blue noise, white noise, and first-order image of the double-slit. The image of the two slits are well separated in the second-order image with blue noise speckle illumination (a), but are indistinguishable in the second-order image with white noise speckle illumination (b), and are completely overlapped in the first-order image (c).The slight asymmetry of the double-slit image distribution in (a) and (b) is present because the iris is not exactly oriented to the middle of the two slits.}
	\label{fig:main_result}
\end{figure*}
The experimental results are shown in Fig.~\ref{fig:main_result}(a) for second-order image with blue noise illumination, Fig.~\ref{fig:main_result}(b) for second-order image with white noise illumination and Fig.~\ref{fig:main_result}(c) for first-order image.   All three measurements are performed under the same conditions, \textit{i.e.}, the same laser power, number of frames, and exposure time of each frame. 
Data are collected by the CMOS sensor with exposure time at 30 $\mu$s for each pattern and averaged over 3000 patterns. It can be clearly seen that, due to the finite pinhole size, the first-order imaging system is not able to resolve the double-slit at all, while the second-order imaging with white noise cannot distinguish the double-slit either and the image is blurred. On the other hand, Fig.~\ref{fig:main_result}(a) shows a clearly distinguished slit image.

\begin{figure*}[hbt!]
\centering\includegraphics[width=0.65\linewidth]{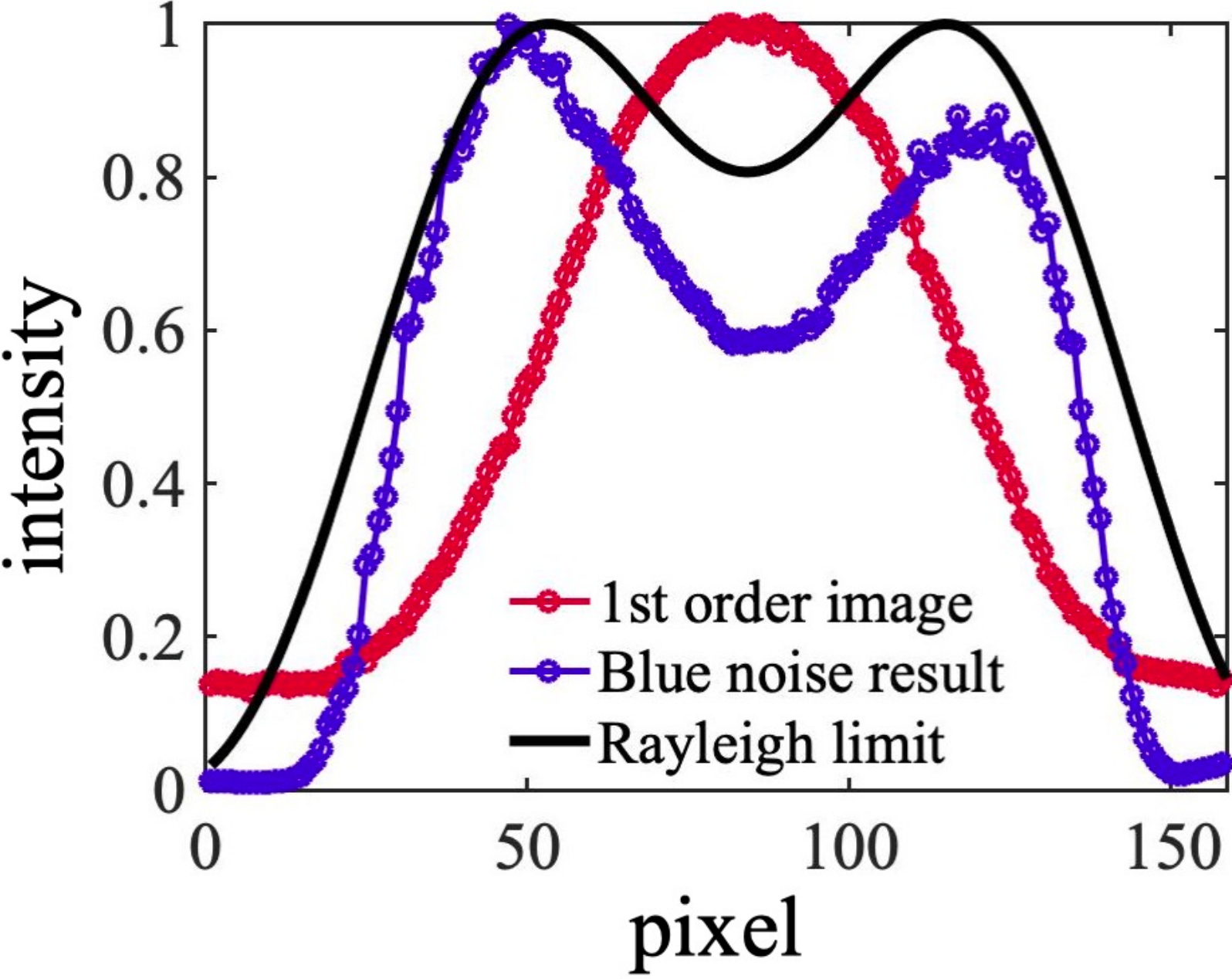}
\caption{The resolution of the first-order and second-order blue noise imaging systems, which is characterized by their corresponding effective lens diameters. 
The solid black line represents the theoretical Rayleigh limit when the effective diameter of the lens is just able to resolve the two slits, which is found to be $2.1$ mm. Red dots are the measured first-order image, fitted with an effective lens diameter of $0.84$ mm. Blue dots are the measured second-order blue noise image, fitted to an effective lens diameter of $2.5$ mm.}
	\label{fig:Rayleighlimit_compare}
\end{figure*}

In Fig.~\ref{fig:Rayleighlimit_compare}, we plot the one-dimensional image of the double-slit when the effective diameter of the lens is just enough to separate the two slits, \textit{i.e.}, at the Rayleigh diffraction limit. The simulation, in solid black lines, shows that a lens with a minimal effective diameter of $\sim$ 2.1 mm is required. We also choose the 1D cross-section of vertical
pixel 55, from the blue noise second-order image and the first-order image results from Fig.~\ref{fig:main_result}. The data are fitted to the first-order imaging equation, providing a quantitative evaluation of the resolving power of the imaging systems. Fitting the first-order image data results in an effective lens diameter of $\sim 0.8$ mm, which is well consistent with our experimental setup of the first-order imaging system. 
The same analysis shows that, to have an equivalent resolution of the second-order blue noise imaging system, one needs to use a lens of $\sim 2.5$ mm effective diameter in the first-order imaging system.  Thus, using the blue noise speckle in second-order correlation imaging, the resolution is enhanced by $\sim 3$ times.

\section{Conclusion}
In summary, we proposed a novel method to synthesize blue noise speckle patterns satisfying Rayleigh statistics by tailoring the amplitudes of the input light fields.
The resulting speckles show a unique feature of negative spatial correlation between neighboring pixels. 
The anti-correlation of spatial intensity fluctuation endows the second-order correlation imaging system a resolving power that is $\sim 3$ times higher than the first-order imaging and is equivalent to a $9$th order correlation imaging system using thermal light illumination.
Our method is versatile and compatible with a broad range of optical setups.
It can be used in computational ghost imaging systems to sharpen the edges of the ghost image and enhance the signal-to-noise ratio.
Our method has potential application in microscopy and biomedical imaging systems such as laser speckle contrast imaging (LSCI) based on the second-order correlation imaging mechanism~\cite{briers1996laser,aminfar2019application,zhang2019laser}. With the buttress of blue boise speckle illumination, LSCI can obtain essential enhancement of resolution.

\section{Conflict of interest}
The authors declare no conflict of interest.

\section{Author contributions}
ZL, TP, MSZ, and MOS initiated the project. ZL and TP proposed the approach, XYN, FY, DYL, XLD and TP did the simulations. ZL, XYN, FY, XPL, XCZ, TP, MSZ and MOS realized the structured noise generation and imaging system.
All authors prepared the manuscript.

\section{Acknowledgements}
The authors thank R. Nessler and H. Xu for helpful discussion, X. N. and F. Y. thank A. Svidzinsky for his help during their visit to IQSE, Texas A\&M University.
We acknowledge financial support by the Air Force Office of Scientific Research (Award No. FA9550-20-1-0366 DEF), Office of Naval Research (Award No. N00014-20-1-2184), Robert A. Welch Foundation (Grant No. A-1261), National Science Foundation (Grant No. PHY-2013771), and Qatar National Research Fund (project NPRP 13S-0205-200258).


\begin{thebibliography}{10}
\expandafter\ifx\csname url\endcsname\relax
  \def\url#1{\texttt{#1}}\fi
\expandafter\ifx\csname urlprefix\endcsname\relax\def\urlprefix{URL }\fi
\providecommand{\bibinfo}[2]{#2}
\providecommand{\eprint}[2][]{\url{#2}}

\bibitem{Abbe:1873aa}
\bibinfo{author}{Abbe, E.}
\newblock \bibinfo{title}{{Beitr{\"a}ge zur Theorie des Mikroskops und der
  mikroskopischen Wahrnehmung}}.
\newblock \emph{\bibinfo{journal}{Archiv f{\"u}r Mikroskopische Anatomie}}
  \textbf{\bibinfo{volume}{9}}, \bibinfo{pages}{413--468}
  (\bibinfo{year}{1873}).

\bibitem{rayleigh1879xxxi}
\bibinfo{author}{Rayleigh, L.}
\newblock \bibinfo{title}{Investigations in optics, with special reference to
  the spectroscope}.
\newblock \emph{\bibinfo{journal}{The London, Edinburgh, and Dublin
  Philosophical Magazine and Journal of Science}} \textbf{\bibinfo{volume}{8}},
  \bibinfo{pages}{261--274} (\bibinfo{year}{1879}).

\bibitem{d2001two}
\bibinfo{author}{D'Angelo, M.}, \bibinfo{author}{Chekhova, M.~V.} \&
  \bibinfo{author}{Shih, Y.}
\newblock \bibinfo{title}{Two-photon diffraction and quantum lithography}.
\newblock \emph{\bibinfo{journal}{Phys. Rev. Lett.}}
  \textbf{\bibinfo{volume}{87}}, \bibinfo{pages}{013602}
  (\bibinfo{year}{2001}).

\bibitem{li2008optical}
\bibinfo{author}{Li, H.} \emph{et~al.}
\newblock \bibinfo{title}{Optical imaging beyond the diffraction limit via dark
  states}.
\newblock \emph{\bibinfo{journal}{Phys. Rev. A}} \textbf{\bibinfo{volume}{78}},
  \bibinfo{pages}{013803} (\bibinfo{year}{2008}).

\bibitem{boto2000quantum}
\bibinfo{author}{Boto, A.~N.} \emph{et~al.}
\newblock \bibinfo{title}{Quantum interferometric optical lithography:
  exploiting entanglement to beat the diffraction limit}.
\newblock \emph{\bibinfo{journal}{Phys. Rev. Lett.}}
  \textbf{\bibinfo{volume}{85}}, \bibinfo{pages}{2733} (\bibinfo{year}{2000}).

\bibitem{gustafsson2000surpassing}
\bibinfo{author}{Gustafsson, M.~G.}
\newblock \bibinfo{title}{Surpassing the lateral resolution limit by a factor
  of two using structured illumination microscopy}.
\newblock \emph{\bibinfo{journal}{J. Microsc.}} \textbf{\bibinfo{volume}{198}},
  \bibinfo{pages}{82--87} (\bibinfo{year}{2000}).

\bibitem{zeng2014nanometer}
\bibinfo{author}{Zeng, X.}, \bibinfo{author}{Al-Amri, M.} \&
  \bibinfo{author}{Zubairy, M.~S.}
\newblock \bibinfo{title}{Nanometer-scale microscopy via graphene plasmons}.
\newblock \emph{\bibinfo{journal}{Phys. Rev. B}} \textbf{\bibinfo{volume}{90}},
  \bibinfo{pages}{235418} (\bibinfo{year}{2014}).

\bibitem{classen2017superresolution}
\bibinfo{author}{Classen, A.}, \bibinfo{author}{von Zanthier, J.},
  \bibinfo{author}{Scully, M.~O.} \& \bibinfo{author}{Agarwal, G.~S.}
\newblock \bibinfo{title}{Superresolution via structured illumination quantum
  correlation microscopy}.
\newblock \emph{\bibinfo{journal}{Optica}} \textbf{\bibinfo{volume}{4}},
  \bibinfo{pages}{580--587} (\bibinfo{year}{2017}).

\bibitem{kiffner2008resonant}
\bibinfo{author}{Kiffner, M.}, \bibinfo{author}{Evers, J.} \&
  \bibinfo{author}{Zubairy, M.}
\newblock \bibinfo{title}{Resonant interferometric lithography beyond the
  diffraction limit}.
\newblock \emph{\bibinfo{journal}{Phys. Rev. Lett.}}
  \textbf{\bibinfo{volume}{100}}, \bibinfo{pages}{073602}
  (\bibinfo{year}{2008}).

\bibitem{liao2010quantum}
\bibinfo{author}{Liao, Z.}, \bibinfo{author}{Al-Amri, M.} \&
  \bibinfo{author}{Zubairy, M.~S.}
\newblock \bibinfo{title}{Quantum lithography beyond the diffraction limit via
  rabi oscillations}.
\newblock \emph{\bibinfo{journal}{Phys. Rev. Lett.}}
  \textbf{\bibinfo{volume}{105}}, \bibinfo{pages}{183601}
  (\bibinfo{year}{2010}).

\bibitem{hell1994breaking}
\bibinfo{author}{Hell, S.~W.} \& \bibinfo{author}{Wichmann, J.}
\newblock \bibinfo{title}{Breaking the diffraction resolution limit by
  stimulated emission: stimulated-emission-depletion fluorescence microscopy}.
\newblock \emph{\bibinfo{journal}{Opt. Lett.}} \textbf{\bibinfo{volume}{19}},
  \bibinfo{pages}{780--782} (\bibinfo{year}{1994}).

\bibitem{hess2006ultra}
\bibinfo{author}{Hess, S.~T.}, \bibinfo{author}{Girirajan, T.~P.} \&
  \bibinfo{author}{Mason, M.~D.}
\newblock \bibinfo{title}{Ultra-high resolution imaging by fluorescence
  photoactivation localization microscopy}.
\newblock \emph{\bibinfo{journal}{Biophys. J.}} \textbf{\bibinfo{volume}{91}},
  \bibinfo{pages}{4258--4272} (\bibinfo{year}{2006}).

\bibitem{rust2006sub}
\bibinfo{author}{Rust, M.~J.}, \bibinfo{author}{Bates, M.} \&
  \bibinfo{author}{Zhuang, X.}
\newblock \bibinfo{title}{Sub-diffraction-limit imaging by stochastic optical
  reconstruction microscopy ({STORM})}.
\newblock \emph{\bibinfo{journal}{Nature Methods}}
  \textbf{\bibinfo{volume}{3}}, \bibinfo{pages}{793--796}
  (\bibinfo{year}{2006}).

\bibitem{betzig2006imaging}
\bibinfo{author}{Betzig, E.} \emph{et~al.}
\newblock \bibinfo{title}{Imaging intracellular fluorescent proteins at
  nanometer resolution}.
\newblock \emph{\bibinfo{journal}{Science}} \textbf{\bibinfo{volume}{313}},
  \bibinfo{pages}{1642--1645} (\bibinfo{year}{2006}).

\bibitem{hemmer2006quantum}
\bibinfo{author}{Hemmer, P.~R.}, \bibinfo{author}{Muthukrishnan, A.},
  \bibinfo{author}{Scully, M.~O.} \& \bibinfo{author}{Zubairy, M.~S.}
\newblock \bibinfo{title}{Quantum lithography with classical light}.
\newblock \emph{\bibinfo{journal}{Phys. Rev. Lett.}}
  \textbf{\bibinfo{volume}{96}}, \bibinfo{pages}{163603}
  (\bibinfo{year}{2006}).

\bibitem{guerrieri2010sub}
\bibinfo{author}{Guerrieri, F.} \emph{et~al.}
\newblock \bibinfo{title}{Sub-{R}ayleigh imaging via $n$-photon detection}.
\newblock \emph{\bibinfo{journal}{Phys. Rev. Lett.}}
  \textbf{\bibinfo{volume}{105}}, \bibinfo{pages}{163602}
  (\bibinfo{year}{2010}).

\bibitem{oh2013sub}
\bibinfo{author}{Oh, J.-E.}, \bibinfo{author}{Cho, Y.-W.},
  \bibinfo{author}{Scarcelli, G.} \& \bibinfo{author}{Kim, Y.-H.}
\newblock \bibinfo{title}{Sub-{R}ayleigh imaging via speckle illumination}.
\newblock \emph{\bibinfo{journal}{Opt. Lett.}} \textbf{\bibinfo{volume}{38}},
  \bibinfo{pages}{682--684} (\bibinfo{year}{2013}).

\bibitem{sprigg2016super}
\bibinfo{author}{Sprigg, J.}, \bibinfo{author}{Peng, T.} \&
  \bibinfo{author}{Shih, Y.}
\newblock \bibinfo{title}{Super-resolution imaging using the spatial-frequency
  filtered intensity fluctuation correlation}.
\newblock \emph{\bibinfo{journal}{Scientific reports}}
  \textbf{\bibinfo{volume}{6}}, \bibinfo{pages}{38077} (\bibinfo{year}{2016}).

\bibitem{chen2017sub}
\bibinfo{author}{Chen, X.-H.}, \bibinfo{author}{Kong, F.-H.},
  \bibinfo{author}{Fu, Q.}, \bibinfo{author}{Meng, S.-Y.} \&
  \bibinfo{author}{Wu, L.-A.}
\newblock \bibinfo{title}{Sub-{R}ayleigh resolution ghost imaging by spatial
  low-pass filtering}.
\newblock \emph{\bibinfo{journal}{Opt. Lett.}} \textbf{\bibinfo{volume}{42}},
  \bibinfo{pages}{5290--5293} (\bibinfo{year}{2017}).

\bibitem{zhou2012resolution}
\bibinfo{author}{Zhou, Y.}, \bibinfo{author}{Liu, J.}, \bibinfo{author}{Simon,
  J.} \& \bibinfo{author}{Shih, Y.}
\newblock \bibinfo{title}{Resolution enhancement of third-order thermal light
  ghost imaging in the photon counting regime}.
\newblock \emph{\bibinfo{journal}{J. Opt. Soc. Am. B}}
  \textbf{\bibinfo{volume}{29}}, \bibinfo{pages}{377--381}
  (\bibinfo{year}{2012}).

\bibitem{oppel2014directional}
\bibinfo{author}{Oppel, S.}, \bibinfo{author}{Wiegner, R.},
  \bibinfo{author}{Agarwal, G.} \& \bibinfo{author}{Von~Zanthier, J.}
\newblock \bibinfo{title}{Directional superradiant emission from statistically
  independent incoherent nonclassical and classical sources}.
\newblock \emph{\bibinfo{journal}{Phys. Rev. Lett.}}
  \textbf{\bibinfo{volume}{113}}, \bibinfo{pages}{263606}
  (\bibinfo{year}{2014}).

\bibitem{zhang2016high}
\bibinfo{author}{Zhang, S.}, \bibinfo{author}{Wang, W.}, \bibinfo{author}{Yu,
  R.} \& \bibinfo{author}{Yang, X.}
\newblock \bibinfo{title}{High-order correlation of non-{R}ayleigh speckle
  fields and its application in super-resolution imaging}.
\newblock \emph{\bibinfo{journal}{Laser Phys.}} \textbf{\bibinfo{volume}{26}},
  \bibinfo{pages}{055007} (\bibinfo{year}{2016}).

\bibitem{martienssen1964coherence}
\bibinfo{author}{Martienssen, W.} \& \bibinfo{author}{Spiller, E.}
\newblock \bibinfo{title}{Coherence and fluctuations in light beams}.
\newblock \emph{\bibinfo{journal}{Am. J. Phys.}} \textbf{\bibinfo{volume}{32}},
  \bibinfo{pages}{919--926} (\bibinfo{year}{1964}).

\bibitem{shapiro2008computational}
\bibinfo{author}{Shapiro, J.~H.}
\newblock \bibinfo{title}{Computational ghost imaging}.
\newblock \emph{\bibinfo{journal}{Phys. Rev. A}} \textbf{\bibinfo{volume}{78}},
  \bibinfo{pages}{061802} (\bibinfo{year}{2008}).

\bibitem{goodman2005introduction}
\bibinfo{author}{Goodman, J.~W.}
\newblock \emph{\bibinfo{title}{Introduction to Fourier optics}}
  (\bibinfo{publisher}{Roberts and Company Publishers}, \bibinfo{year}{2005}).

\bibitem{chen2013100}
\bibinfo{author}{Chen, H.}, \bibinfo{author}{Peng, T.} \&
  \bibinfo{author}{Shih, Y.}
\newblock \bibinfo{title}{100\% correlation of chaotic thermal light}.
\newblock \emph{\bibinfo{journal}{Phys. Rev. A}} \textbf{\bibinfo{volume}{88}},
  \bibinfo{pages}{023808} (\bibinfo{year}{2013}).

\bibitem{li2020photon}
\bibinfo{author}{Li, S.-W.}, \bibinfo{author}{Li, F.}, \bibinfo{author}{Peng,
  T.} \& \bibinfo{author}{Agarwal, G.}
\newblock \bibinfo{title}{Photon statistics of quantum light on scattering from
  rotating ground glass}.
\newblock \emph{\bibinfo{journal}{Phys. Rev. A}}
  \textbf{\bibinfo{volume}{101}}, \bibinfo{pages}{063806}
  (\bibinfo{year}{2020}).

\bibitem{waller2012phase}
\bibinfo{author}{Waller, L.}, \bibinfo{author}{Situ, G.} \&
  \bibinfo{author}{Fleischer, J.~W.}
\newblock \bibinfo{title}{Phase-space measurement and coherence synthesis of
  optical beams}.
\newblock \emph{\bibinfo{journal}{Nature Photonics}}
  \textbf{\bibinfo{volume}{6}}, \bibinfo{pages}{474--479}
  (\bibinfo{year}{2012}).

\bibitem{takasaki2014phase}
\bibinfo{author}{Takasaki, K.~T.} \& \bibinfo{author}{Fleischer, J.~W.}
\newblock \bibinfo{title}{Phase-space measurement for depth-resolved
  memory-effect imaging}.
\newblock \emph{\bibinfo{journal}{Opt. Express}} \textbf{\bibinfo{volume}{22}},
  \bibinfo{pages}{31426--31433} (\bibinfo{year}{2014}).

\bibitem{bromberg2014generating}
\bibinfo{author}{Bromberg, Y.} \& \bibinfo{author}{Cao, H.}
\newblock \bibinfo{title}{Generating non-{R}ayleigh speckles with tailored
  intensity statistics}.
\newblock \emph{\bibinfo{journal}{Phys. Rev. Lett.}}
  \textbf{\bibinfo{volume}{112}}, \bibinfo{pages}{213904}
  (\bibinfo{year}{2014}).

\bibitem{bender2018customizing}
\bibinfo{author}{Bender, N.}, \bibinfo{author}{Y{\i}lmaz, H.},
  \bibinfo{author}{Bromberg, Y.} \& \bibinfo{author}{Cao, H.}
\newblock \bibinfo{title}{Customizing speckle intensity statistics}.
\newblock \emph{\bibinfo{journal}{Optica}} \textbf{\bibinfo{volume}{5}},
  \bibinfo{pages}{595--600} (\bibinfo{year}{2018}).

\bibitem{kondakci2016sub}
\bibinfo{author}{Kondakci, H.~E.}, \bibinfo{author}{Szameit, A.},
  \bibinfo{author}{Abouraddy, A.~F.}, \bibinfo{author}{Christodoulides, D.~N.}
  \& \bibinfo{author}{Saleh, B.~E.}
\newblock \bibinfo{title}{Sub-thermal to super-thermal light statistics from a
  disordered lattice via deterministic control of excitation symmetry}.
\newblock \emph{\bibinfo{journal}{Optica}} \textbf{\bibinfo{volume}{3}},
  \bibinfo{pages}{477--482} (\bibinfo{year}{2016}).

\bibitem{li2016generation}
\bibinfo{author}{Li, X.} \emph{et~al.}
\newblock \bibinfo{title}{Generation of a super-{R}ayleigh speckle field via a
  spatial light modulator}.
\newblock \emph{\bibinfo{journal}{Appl. Phys. B}}
  \textbf{\bibinfo{volume}{122}}, \bibinfo{pages}{82} (\bibinfo{year}{2016}).

\bibitem{zhou2017superbunching}
\bibinfo{author}{Zhou, Y.} \emph{et~al.}
\newblock \bibinfo{title}{Superbunching pseudothermal light}.
\newblock \emph{\bibinfo{journal}{Phys. Rev. A}} \textbf{\bibinfo{volume}{95}},
  \bibinfo{pages}{053809} (\bibinfo{year}{2017}).

\bibitem{ulichney1988dithering}
\bibinfo{author}{Ulichney, R.~A.}
\newblock \bibinfo{title}{Dithering with blue noise}.
\newblock \emph{\bibinfo{journal}{Proc. IEEE}} \textbf{\bibinfo{volume}{76}},
  \bibinfo{pages}{56--79} (\bibinfo{year}{1988}).

\bibitem{Bender:19}
\bibinfo{author}{Bender, N.}, \bibinfo{author}{Y{i}lmaz, H.},
  \bibinfo{author}{Bromberg, Y.} \& \bibinfo{author}{Cao, H.}
\newblock \bibinfo{title}{Introducing non-local correlations into laser
  speckles}.
\newblock \emph{\bibinfo{journal}{Opt. Express}} \textbf{\bibinfo{volume}{27}},
  \bibinfo{pages}{6057} (\bibinfo{year}{2019}).

\bibitem{shih2018introduction}
\bibinfo{author}{Shih, Y.}
\newblock \emph{\bibinfo{title}{An introduction to quantum optics: photon and
  biphoton physics}} (\bibinfo{publisher}{CRC press}, \bibinfo{year}{2018}).

\bibitem{briers1996laser}
\bibinfo{author}{Briers, J.~D.} \& \bibinfo{author}{Webster, S.}
\newblock \bibinfo{title}{Laser speckle contrast analysis (lasca): a
  nonscanning, full-field technique for monitoring capillary blood flow}.
\newblock \emph{\bibinfo{journal}{J. Biomed. Opt.}}
  \textbf{\bibinfo{volume}{1}}, \bibinfo{pages}{174--180}
  (\bibinfo{year}{1996}).

\bibitem{aminfar2019application}
\bibinfo{author}{Aminfar, A.}, \bibinfo{author}{Davoodzadeh, N.},
  \bibinfo{author}{Aguilar, G.} \& \bibinfo{author}{Princevac, M.}
\newblock \bibinfo{title}{Application of optical flow algorithms to laser
  speckle imaging}.
\newblock \emph{\bibinfo{journal}{Microvascular Research}}
  \textbf{\bibinfo{volume}{122}}, \bibinfo{pages}{52--59}
  (\bibinfo{year}{2019}).

\bibitem{zhang2019laser}
\bibinfo{author}{Zhang, R.} \emph{et~al.}
\newblock \bibinfo{title}{Laser speckle imaging for blood flow based on pixel
  resolved zero-padding auto-correlation coefficient distribution}.
\newblock \emph{\bibinfo{journal}{Opt. Commun.}}
  \textbf{\bibinfo{volume}{439}}, \bibinfo{pages}{38--46}
  (\bibinfo{year}{2019}).

\end{thebibliography}

\end{document}